\documentclass[preprint,12pt]{elsarticle}

\usepackage{amssymb,color}
\newcommand{\black}{\color{black}}
\newcommand{\red}{\color{red}}

\newcommand{\green}{\color{green}}

\renewcommand{\red}{\black}
\renewcommand{\green}{\black}

\journal{Astroparticle Physics}

\begin{document}

\begin{frontmatter}



\title{No muon excess in extensive air showers at \red
100--500~PeV primary energy:
EAS--MSU results
\black
}


\author[a]{Yu.\,A.\,Fomin}
\author[a]{N.\,N.\,Kalmykov}
\author[b]{I.\,S.\,Karpikov}
\author[a]{G.\,V.\,Kulikov}
\author[b]{M.\,Yu.\,Kuznetsov}
\author[b]{G.\,I.\,Rubtsov}
\author[a]{V.\,P.\,Sulakov}
\author[b]{S.\,V.\,Troitsky}
\ead{st@ms2.inr.ac.ru}

\address[a]{D.V.~Skobeltsyn Institute of Nuclear Physics, \\
M.V.~Lomonosov Moscow State University, Moscow 119991, Russia}
\address[b]{Institute for Nuclear Research of the Russian Academy of
Sciences,\\
60th October Anniversary prospect 7A, 117312 Moscow, Russia}

\begin{abstract}
Some discrepancies
have been reported between observed and simulated muon content of
extensive air showers: the number of observed muons exceeded the
expectations in HiRes-MIA, Yakutsk and Pierre Auger Observatory data. Here,
we analyze the data of the Moscow State University Extensive Air Shower
(EAS--MSU) array on $E_{\mu} \gtrsim 10$~GeV muons in showers caused by
$\sim (10^{17}-10^{18})$~eV primary particles and demonstrate that they
agree with simulations (QGSJET-II-04 hadronic interaction model) once the
primary composition inferred from the surface-detector data is assumed.
\end{abstract}


\end{frontmatter}


\section{Introduction}
\label{sec:intro}
Ultra-high-energy cosmic rays provide 
a unique laboratory to study
hadronic interactions at the center-of-mass energies and in kinematical
regimes not accessible at colliders. Modelling of the development of an
extensive air shower (EAS), a cascade process in the terrestrial
atmosphere initiated by an energetic cosmic particle, requires
an extrapolation of verified interaction models. Not surprisingly, this
often results in discrepancies between measured and simulated EAS
properties, or between physical properties of the primary particle
reconstructed by different methods. A well-known result,
\red
possibly related to the lack of understanding of the EAS development
\cite{New-PAO},
\black
is the systematic difference between the primary energies $E$
reconstructed by the fluorescen\red{}ce\black \ detectors and by surface
arrays for very same events, as seen by the Pierre Auger Observatory (PAO)
\cite{Engel} and the Telescope Array (TA) experiment \cite{TA-E}. It may
or may not be related to the apparent excess of muons ($E_{\mu }
\gtrsim$GeV\red)\black \ in EAS reported at $E \gtrsim 10^{19}$~eV by the
PAO \green \cite{New-PAO, Engel, PAOmu2} \black and Yakutsk \cite{Yak-mu}
experiments. A similar excess had been observed earlier by the HiRes/MIA
experiment at $E \gtrsim 10^{17}$~eV~\cite{HiRes-MIA}. The purpose of the
present study is to compare observed and simulated densities of $E_{\mu
}>10$~GeV muons in air showers induced by $E\sim 10^{17}$~eV primaries,
based on the EAS-MSU data.

A subtle point of all comparisons of this kind is that the muon content of
a EAS depends strongly on the type of a primary particle. As a result, the
average muon content in the MC set depends not only on the hadronic
interaction model used, but also on the primary composition assumed at
the simulation. Therefore, for a meaningful comparison, one needs an
independent estimator of the primary composition in the very same data set
for which the muon data are ana\green{l}\black{}ysed. An estimator of this
kind is often missing. In this work, we take advantage of the knowledge of
the primary composition obtained from the surface-detector data only, as
discussed below.

The rest of the paper is organized as follows.
In Sec.~\ref{sec:data}, a brief description of the installation and of the
data set is given, together with references to previous more detailed
publications \red about the EAS-MSU array\black. In Sec.~\ref{sec:anal}, we
discuss the analysis performed in this work. Section~\ref{sec:results}
presents our results, while Sec.~\ref{sec:concl} summarizes our
conclusions.

\section{Data}
\label{sec:data}
\paragraph{Installation}
The EAS-MSU array \cite{EAS-MSU}, located in Moscow \red (190~m a.s.l.,
corresponding to the atmospheric depth $X\approx 990$~g/cm$^2$)\black,
operated \red in 1955--\black{}1990 in various configurations. A detailed
description of the array in the ultimate configuration, whose data are
discussed here, may be found in Ref.~\cite{JINST}. The total area of the
array, 0.5~km$^{2}$, was covered with 76 charged-particle detector
stations, each consisting of multiple Geiger--Mueller counters. A unique
feature of the installation was the presence of large-area muon detectors.
The main muon detector, located in the array  center, had \red a \black
total area of 36.4~m$^{2}$ and consisted of \red 1104 \black similar
Geiger-Mueller counters, each with \red an \black area of 0.033~m$^{2}$,
located at \red a \black depth of 40 meters of water equivalent
underground. It is the data of this detector \red what \black we study
here.

The geometry of the array, the trigger system and the reconstruction
procedure are described in detail in Ref.~\cite{JINST}. The surface
detector stations allow to reconstruct the lateral distribution function
(LDF) for charged particles, parametrized as
\begin{equation}
\rho(S,  r  )=
N_{e}C(S)(  r  /R_{0})^{(S+\alpha(  r  )-2)}\cdot(  r
/R_{0}+1)^{(S+\alpha(  r  )-4.5)},
\label{Eq:LDF}
\end{equation}
where $\rho$ is the particle density \red at the detector plane (we study
vertical showers, $\theta \le 30^{\circ}$, and thus neglect the
asymmetry caused by attenuation effects)\black, $r$ is the distance to the
shower axis, $R_{0}= 80$~m is the Moliere radius, $S$ is the modified age
parameter of the EAS, $C(S)$ is the normalization coefficient\red, which
is \black calculated numerically{\green,} and $\alpha( r  )$ is the
correction to $S$ determined empirically and presented e.g.\ in
Ref.~\cite{JINST}, $0 \lesssim \alpha \lesssim 0.4$\red; the normalization
$N_{e}$ determines the total number of charged particles.\black

\paragraph{Basic selection cuts}
The following cuts were imposed for the high-energy surface-detector data
sample studied in this work:

1. Convergence of the reconstruction.

2. The LDF age parameter \red belongs to the interval \black $0.3<S<1.8$.

3. The reconstructed zenith angle \red satisfies \black $\theta<30^\circ$.

4. The reconstructed shower axis is within 240~m from the array
center\green, where the muon detector is located\black.

5. The reconstructed shower size is $N_{e}>2\times10^7$, which
corresponds to $E \gtrsim \green  10^{16.5}\black$~eV, \green
assuming protons as primaries \black.

For this study, we use the data recorded in the period 1984 --- 1990
(1372 days). After application of the cuts, the surface-detector data sample
contains 922 events. \red This sample is representative in the sense
that the event set agrees with the Monte-Carlo
simulations based on the EAS-MSU spectrum, as discussed in
Ref.~\cite{JINST}\black\green. The efficiency of the installation is
greater than 95\% for primary energies $E \gtrsim 10^{16.5}$~eV for these
cuts. \black

\paragraph{\black Muon detectors}
\green
As it has been mentioned above, the muon detectors are shielded by a thick
layer of soil which effectively absorbs all the shower particles but
muons: the fraction of other particles in the detector is far below the
$10^{-3}$ level. \black The threshold energy of vertical muons registered
by \red the \black underground detectors \cite{JINST}
is\green\footnote{\green As it has been discussed in
Ref.~\cite{JINST}, the detector is shielded by the soil of
40~m.w.e. The energy-dependent muon absorption is treated in the
continuous-slowing-down approximation \cite{CSDA}. For zenith
angles studied here ($\theta \le 30^{\circ}$), the threshold
energy scales obviously with $\cos\theta$.}\black $E_{\mu }\approx 10$~GeV.
The main muon detector consisted of 32 independent sections. The muon
density is estimated as $\rho_\mu = \ln \left(n/(n-m)\right)/A $, where
$n$ is the number of counters in the muon detector, $m$ is the number of
fired counters and $A$ is the area of the counter\red\footnote{\red This
formula follows from the binomial distribution. Indeed, the probability
that one counter does not fire is estimated as $\exp \left( - \rho_{\mu} A
\right)$. The probability to have $m$ counters fired \red is
$$
P(m,p)=\frac{n!}{m!(n-m)!}p^m (1-p)^{n-m},
$$
where $p = 1 - \exp \left(- \rho_{\mu} A \right)$.
It is maximized for \green $p=m/n$\black, which results in $\exp
\left( - \rho_{\mu} A \right) =(n-m)/n$.}. In our analysis, we further
restrict the data sample to the days when not less than 28 of 32 sections
were operational \red and no failure of the muon detector was recorded in
24 hours\black. This removes \red 168 \black days of data, so the final
sample for muon studies contains \red 809 \black events \red recorded
during the live time of 14060 hours, so that the overall exposure is $7.71
\times 10^{6}$~km~s~sr\black.

\section{Analysis}
\label{sec:anal}
\begin{figure}
\centerline{\includegraphics[width=0.67\columnwidth]{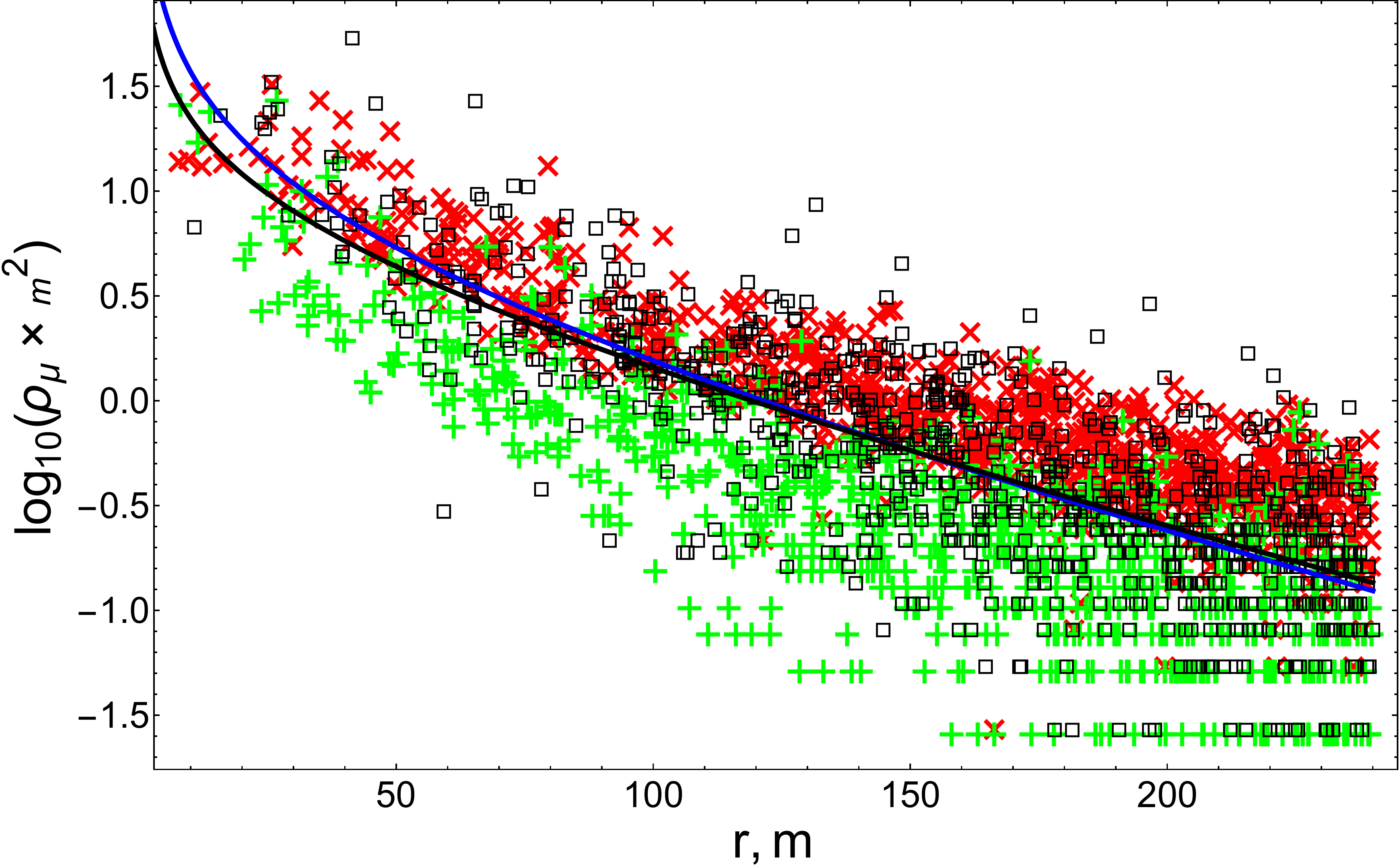}}
\caption{\label{fig:muLDF}
The muon LDF. Points denote individual detector readings (black open
boxes: data; red crosses: MC iron; green pluses: MC protons) for the whole
selected collection of events described in the paper. Black line --
Eq.~(\ref{Eq:muLDF}); blue line -- the best fit described in the text.
\green $N_{e}>2\times10^{7}$, $\theta<30^{\circ}$.
Individual error bars suppressed for clarity. \black }
\end{figure}

The muon density in an air shower decreases with the distance from the
axis\red,\black \ $r$, so to compare this quantity between data and MC, one
often uses the LDF to recalculate the density to a particular fixed value
of $r$. In this study, we use $\rho_{\mu}(100)$, the muon surface density
recalculated to \red the typical core-detector distance of \black $r=100$~m
with the help of the EAS-MSU muon LDF determined in previous works
\cite{EAS-MSU-muon-LDF},
\begin{equation}
\rho_\mu (r) = N_\mu \left(\frac{r}{R_0} \right)^{-\red a_{\mu}\black}
\exp
\left(-r/R_0 \right)\red,
\label{Eq:muLDF}
\end{equation}
where $R_{0}$ is fixed as in Eq.~(\ref{Eq:LDF}) and $a_{\mu}=0.7$.
This muon LDF agrees reasonably well with the data as well as with MC
simulations. Figure~\ref{fig:muLDF}
compares the LDF used in our study with simulations performed for proton
and iron primaries. Allowing for variations of the $a_{\mu }$ parameter in
Eq.~(\ref{Eq:muLDF}), we obtain $a_{\mu }=0.77 \pm 0.02$ for primary
protons, $a_{\mu }=0.54 \pm 0.02$ for primary iron and $\green a_{\mu
}=0.60 \pm 0.13$ \black for the best-fit mixture (57\% Fe, 43\% p)
describing the surface-detector data, see below. The same fitting
procedure for the data gives $\green a_{\mu }=0.74 \pm 0.21$\black. \green
Data and simulations agree well with the value $a_{\mu}=0.7$ fixed in the
analysis. \black

\begin{figure}
\centerline{\includegraphics[width=0.67\columnwidth]{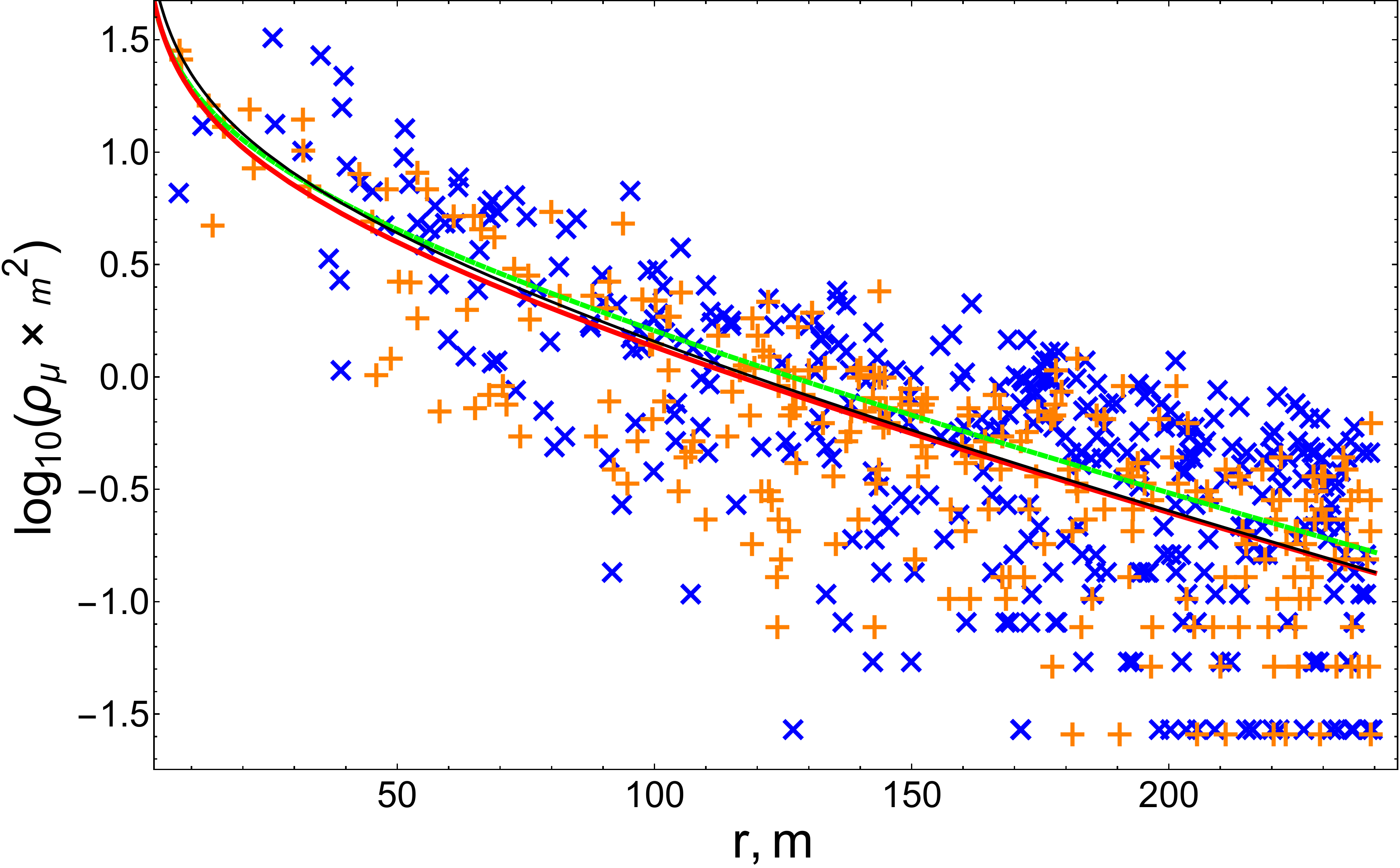}}
\caption{\label{fig:muLDFtheta}
The muon LDF for different zenith angles. Points denote individual
detector readings for MC mixture of 57\% iron and 43\% protons (orange
pluses: $\theta<10^{\circ}$; blue crosses: $25^{\circ}< \theta
<30^{\circ}$) for the whole
selected collection of events described in the paper. Black line -- Eq.~(\ref{Eq:muLDF}); red and green lines --
the best fits for the two $\theta$ ranges.
\green $N_{e}>2\times10^{7}$.
Individual error bars suppressed for clarity.
\black}
\end{figure}

The limited number of events in the data set and the limited range used
make the experimental study of the possible $\theta$ dependence in the
muon LDF hardly possible. In Fig.~\ref{fig:muLDFtheta},
we compare Monte-Carlo simulations \green (QGSJET-II-04) \black for
vertical ($\theta< 10^{\circ}$) and the most inclined ($25^{\circ}<\theta
<30^{\circ}$) events. \green Separate fits for these $\theta$ ranges
result \black in $a_{\mu }=0.65 \pm 0.16$ (vertical) and $a_{\mu} = 0.60
\pm 0.18$ (inclined), \green which \black supports the use of
$\theta$-independent muon LDF in our study.

\black
Comparison of the distributions in $\rho_{\mu}(100)$ for the real and
simulated data constitutes the main part of this work. To produce a
reliable simulation, we make use of the full Monte-Carlo (MC) model of the
installation. This model, developed and described in detail in
Ref.~\cite{JINST}, accounts for the air-shower \red production
and \black development in the atmosphere, its detection and reconstruction.
\red
The air-shower simulations have been performed with the CORSIKA 7.400.1
code \cite{CORSIKA}, using QGSJET-II-04 \cite{QGSJET-II-04} as the
high-energy hadronic interaction model, FLUKA 2011.2c \cite{FLUKA} as the
low-energy hadronic interaction model and EGS-4 \cite{EGS-4} as the
electromagnetic-interaction model. No thinning was employed; the Central
European atmosphere for October 14, 1993 (CORSIKA model number 7) was
used\red\footnote{\red The mean atmospheric pressure in Moscow
for 1984--1991, recalculated to the sea level, is $(1015.9 \pm
5.8)$~hPa \cite{yak}. Correction to the installation altitude of
190~m gives $(989.6 \pm 5.6)$~hPa. The CORSIKA model 7 gives
1020.1~hPa for the sea level and 997.7~hPa for 190~m.}. The detector
response to individual particles was modelled as described in
Ref.~\cite{JINST}. \black
\green
We used 1370 independent CORSIKA showers (852 protons and 518 iron
nuclei), from which we derived, by random moving of the shower axis, our
MC sample (see Ref.~\cite{JINST} for more details). After applying all
cuts, the MC sample contains 4468 proton-induced and 1093 iron-induced
showers. \black

The artificial events are recorded in the same data format as the real
ones and are processed with the same reconstruction software. \red This
means that all selection effects related to the trigger, threshold and
efficiency are accounteed for in the simulation in the same way as in the
real data. \black The resulting distributions of the reconstructed
surface-detector parameters agree well between data and MC \cite{JINST}.
\red The estimated uncertainty (68\% containment radius around the mean)
in the reconstruction of the axis position is 5.7~m; for the arrival
direction it is 1.1$^{\circ}$ \green and \black for $N_{e}$ it is 16.5\%.
\black

The muon density in air showers is a composition-dependent observable:
heavier primary nuclei produce more muons. Many surface-detector
observables are degenerate with respect to the primary composition, and a
good description of \red these \black data might, in principle, be achieved
for various assumptions about the composition. That would lead to quite
different predictions for the muon content, however. To perform a detailed
comparison of muon data with simulations, one therefore requires the
knowledge of the primary composition fully independent on muon detector
readings. This is not always easy to achieve, and even in modern
experiments, surface-detector composition studies are complicated and not
very precise, see e.g.\ Refs.~\cite{PAO-SD-comp, TA-SD-comp}. Fortunately,
the EAS-MSU setup provided for a required observable. Its surface array
was very dense in its central part, and the slope of the LDF, parametrized
by the age parameter $S$ in Eq.~(\ref{Eq:LDF}), is determined with a
precision sufficient to
\red determine the average primary composition with a reasonable accuracy
\black \cite{JINST}. Indeed, an event in the sample we consider has, on
average, 15 detectors used for the LDF reconstruction, which guarantees
the required accuracy.
\red We note that\green, in principle,\black the LDF slope may also, like
the muon number, demonstrate disagreement between data and simulations
with present-day interaction models\green. T\black{}his fact, however, is
not very relevant for the present study, since most of the events \green
lay \black on the distance not exceeding few hundred meters from the center
of the array\green.
The MC and measured age parameters are in
agreement in this radial distance range.\black

\black

In Ref.~\cite{JINST}, working in the frameworks of the two-component
mixture of primary particles (protons and iron), we determined\red, for
$N_{e}>2\times 10^{7}$, \black the best-fit composition which describes the
surface-detector data, including the $S$ distribution. \red Within the
statistical uncertainties, the composition does not change with energy in
the narrow range of energies we study. \black We use this mixture (43\%
protons and 57\% iron\red, following the common power-law spectrum
$\propto E^{-3.1}$\black) in MC simulations and obtain the expected
distribution in $\rho_{\mu} (100)$ which we compare to that observed in
the real data. Then, in order to quantify potential muon excess in data
over simulations, we introduce the coefficient $k$ by which the muon
number is scaled in simulated showers. By definition, $k=1$ corresponds to
the muon number predicted in our simulations with the QGSJET-II-04
hadronic interaction model and with the determined above primary
composition. By means of the binned \red chi-squared test\black, we compare
$\rho_{\mu} (100)$ distributions derived for various $k$ with the data and
determine, consequently, the allowed range of $k$. The scaling of the muon
number was implemented only for muon density measured by underground
detectors, but not for the surface-detector observables. We discuss the
justification for this assumption in Appendix.

\section{Results}
\label{sec:results}
Figure~\ref{fig:rho_mu_dis_S}
\begin{figure}
\centerline{\includegraphics[width=0.67\columnwidth]{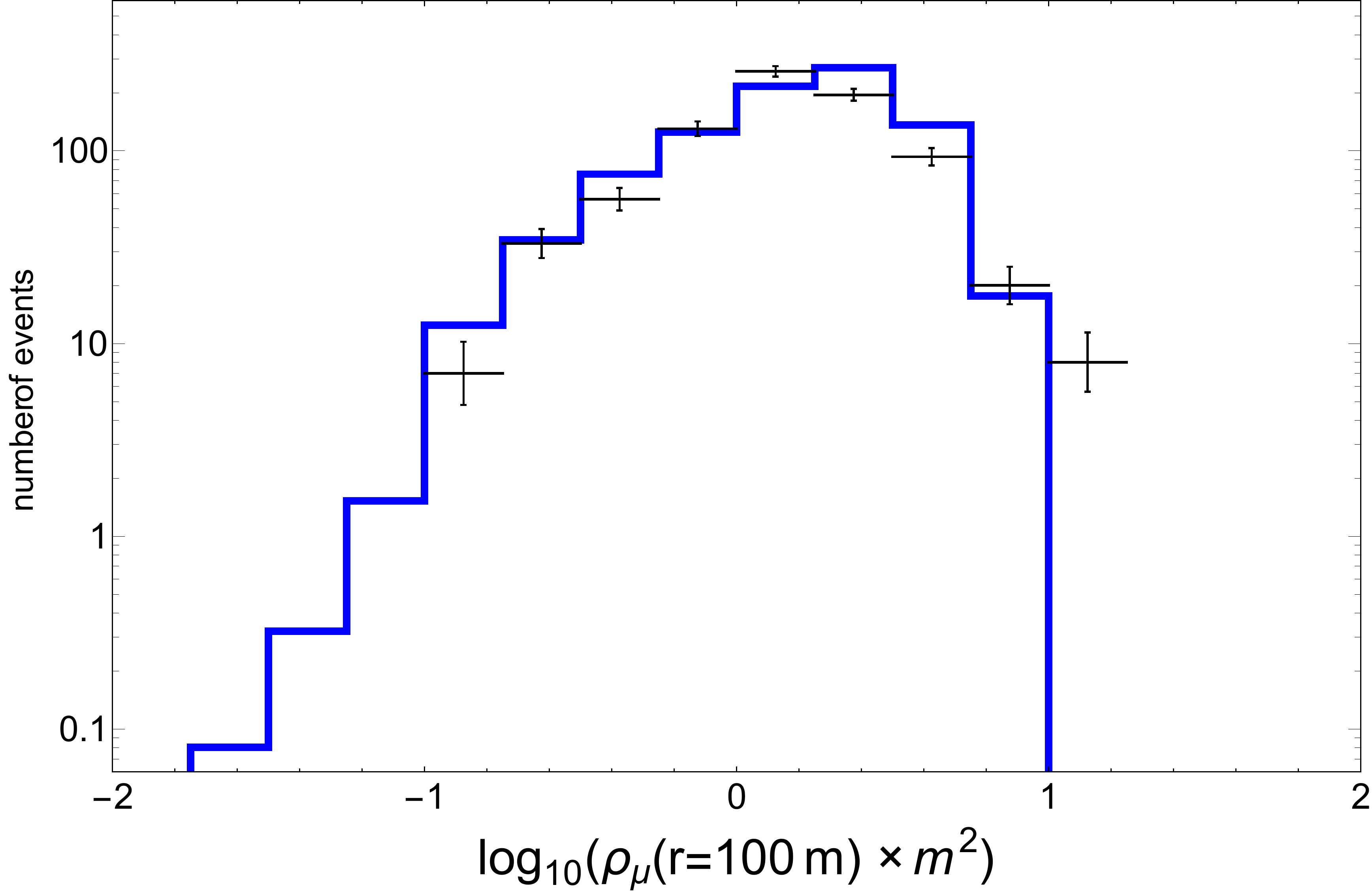}}
\caption{\label{fig:rho_mu_dis_S}
Distribution of $\rho_{\mu}(100)$ in the data sample \red ($N_{e}>2
\times 10^{7}$)\black. Points with error bars \green (statistical errors
only)\black: data; blue histogram: MC simulations based on the primary
composition inferred from the surface-detector data (43\% protons and 57\%
iron), QGSJET-II-04.}
\end{figure}%
compares the distributions in $\rho_{\mu}(100)$ obtained from the
simulations described above ($k=1$) and from the data.
One may see
that the distributions are in a good agreement.

\begin{figure}
\begin{center}
\includegraphics[width=0.67\columnwidth]{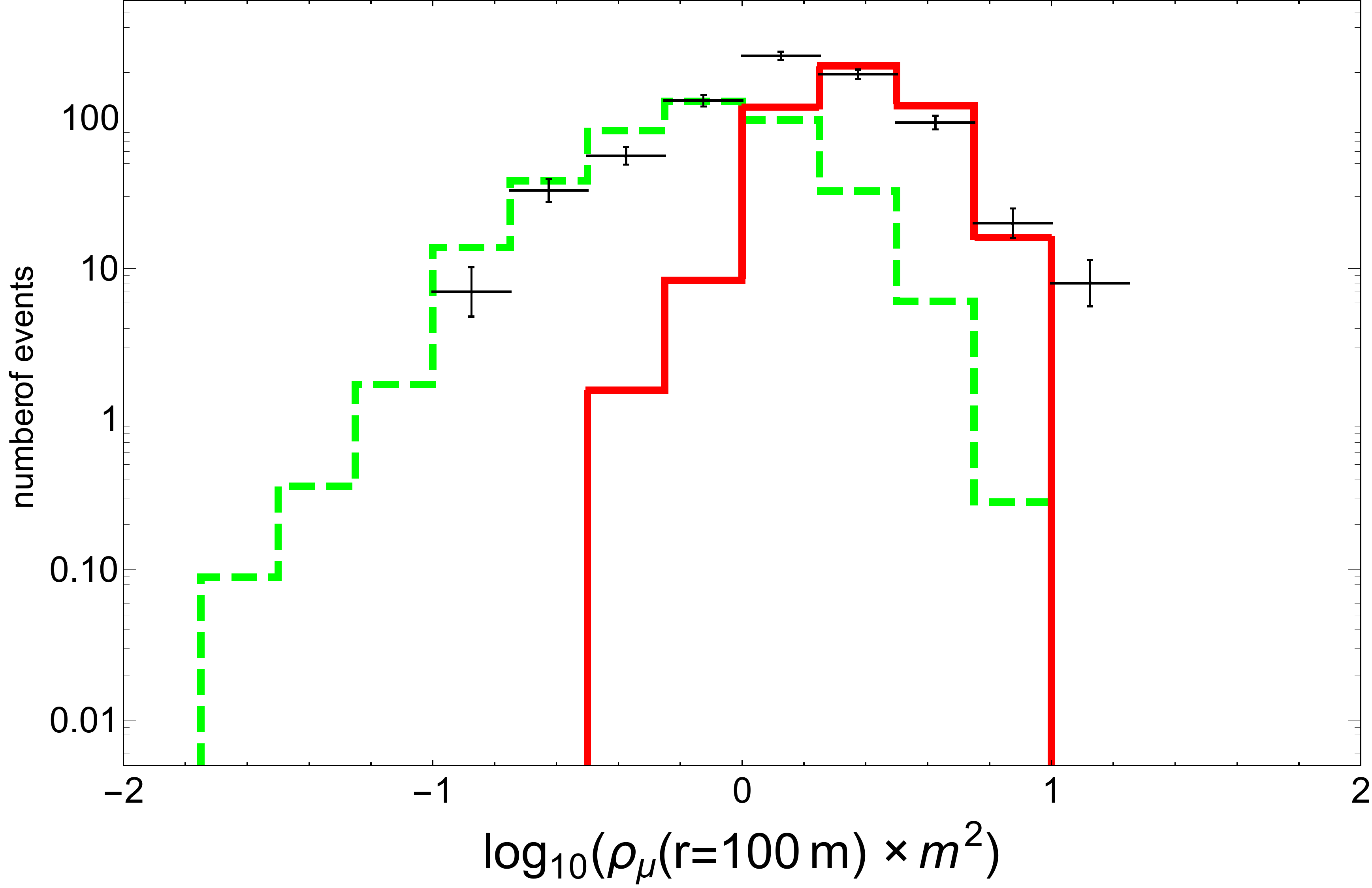}\\
(a)\\  ~\\
\includegraphics[width=0.67\columnwidth]{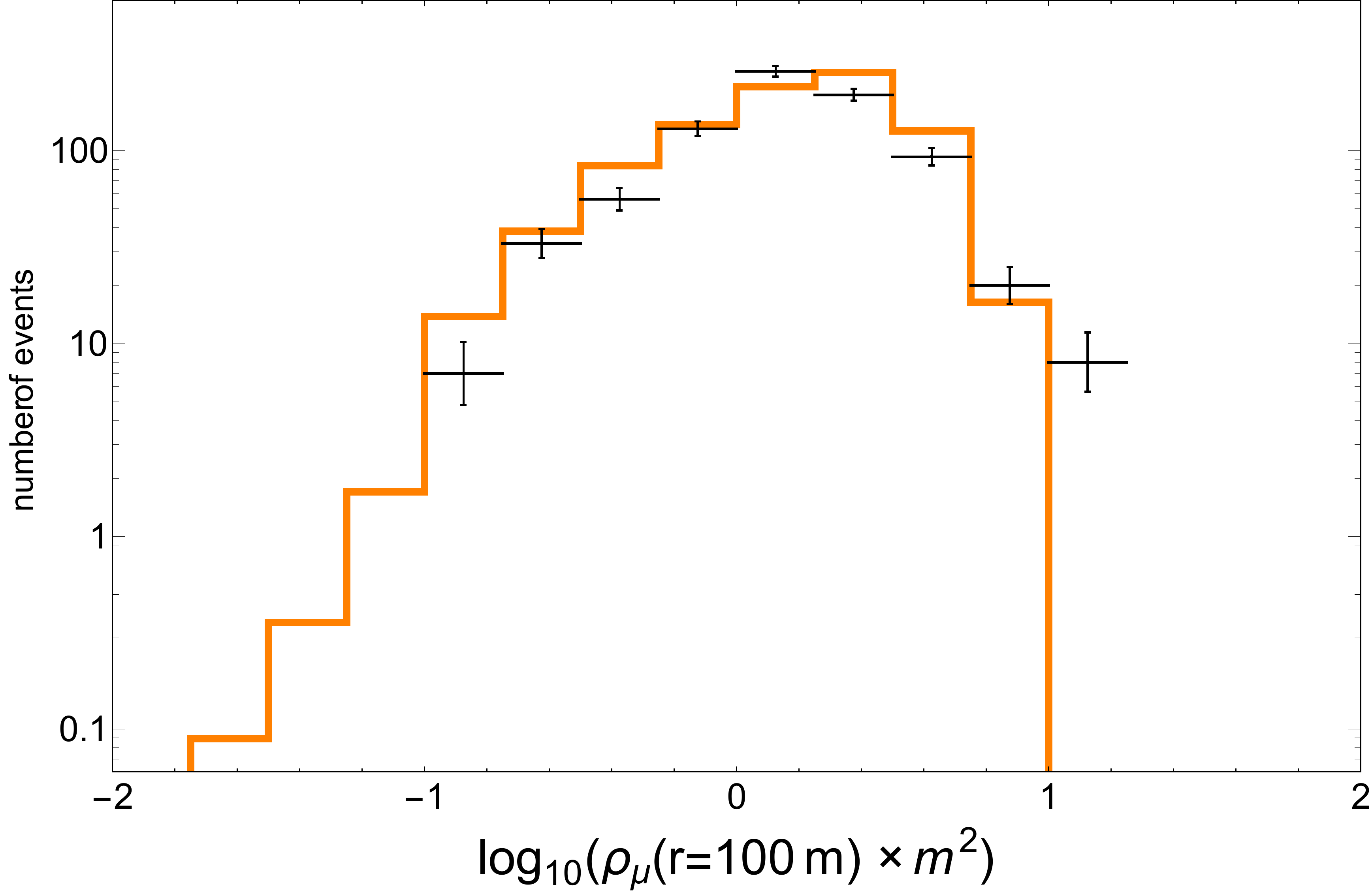}\\
(b)\\
\end{center}
\caption{\label{fig:rho_mu_dis_mu}
Distribution of $\rho_{\mu}(100)$ in the data sample.
Points with error bars \green (statistical errors only)\black: data.
(a) green dashed histogram: MC, protons; red full histogram: MC, iron;
(b) orange histogram: MC,
best-fit primary composition inferred from these \red $\rho_{\mu}(100)$
distributions (46\% protons and 54\% iron)\black.}
\end{figure}

To further quantify this agreement, one may proceed in two ways. Firstly,
note that the muon content depends strongly on the assumed primary
composition, see Fig.~\ref{fig:rho_mu_dis_mu}(a).
We find \red that \black the proton/iron mixture which describes the
$\rho_{\mu}(100)$ distribution \red is  $(54 \pm 6)\%$ \black iron, see
Fig.~\ref{fig:rho_mu_dis_mu}(b) for the best \red binned-likelihood
\black fit, which agrees with 57\% iron determined from the
surface-detector data.

\begin{figure}
\centerline{\includegraphics[width=0.67\columnwidth]{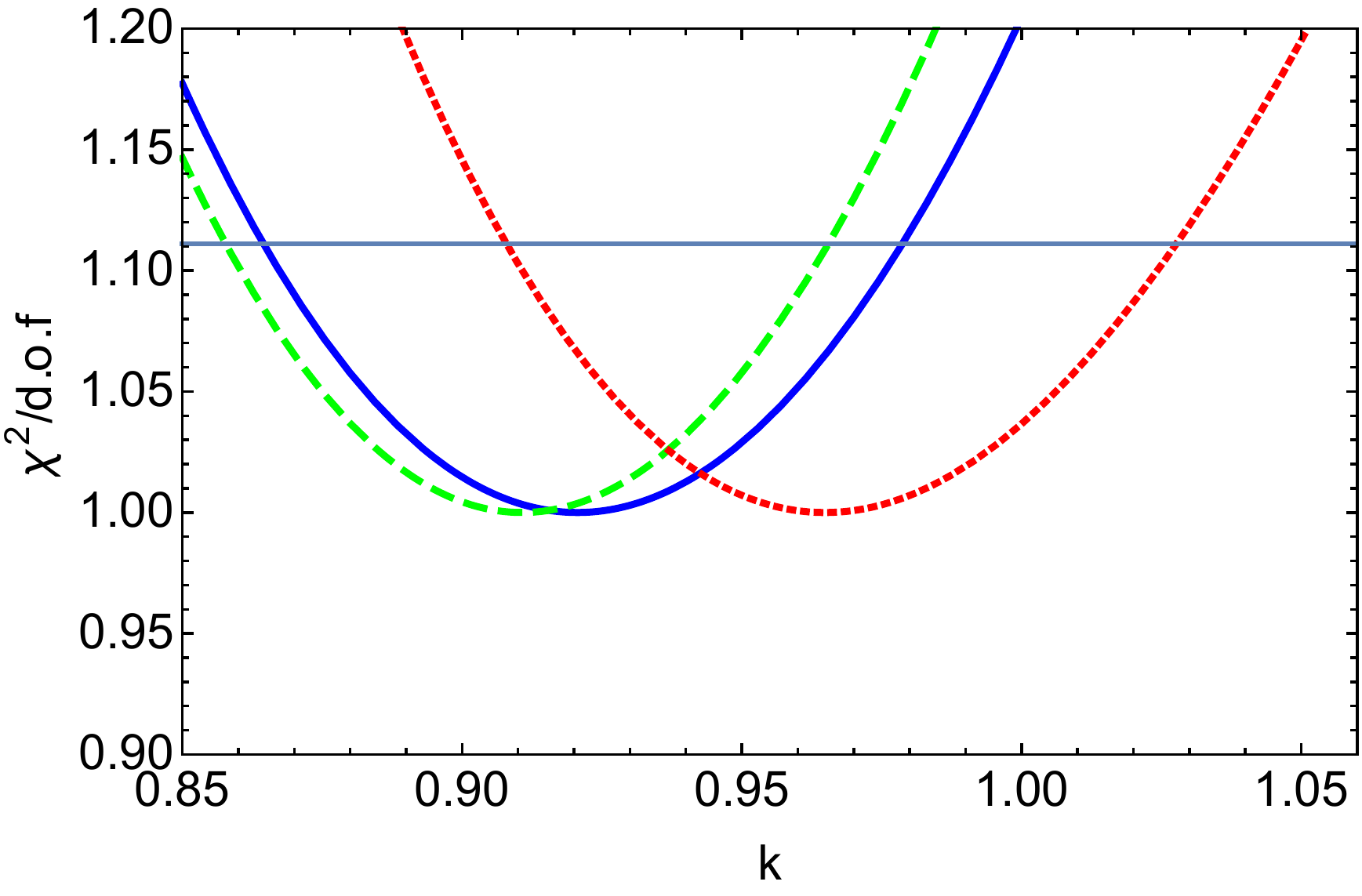}}
\caption{\label{fig:chi2}
The \red $\chi^{2}$ per degree of freedom \black for the muon enhancement
coefficient $k$ determined in the text. Blue \red continuous \black line:
\red assuming \black the EAS-MSU surface-detector composition; green
dashed line: the KASCADE-Grande composition \cite{KASCADE-comp}; red
dotted line: the Tunka-133 composition \cite{Tunka-comp}. \red The
horizontal line represents 68\% CL. \black }
\end{figure}

Secondly, we study how the scaling of the muon number, that is the
variation of $k$, affects the agreement between data and simulations for
the $\rho_{\mu}(100)$ distribution. This comparison is illustrated in
Fig.~\ref{fig:chi2},
where the normalized \red chi-squared \black is presented as a function of
$k$. The standard statistical analysis results in $k=\red 0.92 \pm
0.06$\black, so that no muon excess in data is observed and \red $k>1$
is excluded by the data at the 92\% CL.
\black

\red
All statistical uncertainties \green (related to air-shower fluctuations
and fluctuations in the detector response, as well as uncertainties in the
reconstruction procedure) together with \black potential reconstruction
biases are taken into account by the procedure we use in which data and MC
are processed in the same way. Intrinsic fluctuations of the muon density
are taken into account because we use no-thinning simulations. The
uncertainty in the determination of the muon density is taken into account
in simulations, since the probability of the counter to fire being hit by
a muon is close to $100\%$. However, there are potential sources of
systematic uncertainties which we are turning to now.

\paragraph{Muon LDF} The systematic uncertainty, introduced by the
variations of $a_{\mu}$ to the Fe fraction determined from $\rho_{\mu}$,
does not exceed $\pm1\%$.

\paragraph{Composition} \black
The main result of the paper is based on the primary composition
derived from surface-detector data for the same data set. We tested
its stability \red when \black relaxing this assumption. Assuming the
p/Fe mixture \red reproducing the $\langle \ln A \rangle$ results of \black
KASCADE-Grande \cite{KASCADE-comp} (\red Fig.~7 \green of
Ref.~\cite{KASCADE-comp}\black, stars; \black 59\% iron), we obtain
$k=0.9\red1 \black \pm 0.0\red5\black$; for Tunka-133~\cite{Tunka-comp}
(51\% iron) it is $k=0.9\red6\black \pm 0.0\red6\black$. \red The
confidence levels of exclusion of $k>1$ are 95\% and 67\%, respectively.

One may note that, suggested by results of other experiments, the
composition at $E\sim 10^{17}$~eV changes due to the ``heavy knee'' in the
spectrum. However, working within a rather narrow energy interval, we do
not have sufficient statistics to trace this change in the data\green.
\black We also do not go beyond the two-component proton-iron mixture.
Previous studies have demonstrated \cite{JPG} that the EAS-MSU experiment
does not have enough statistics to distinguish the contribution of
medium-weight nuclei. On the other hand working with the p-Fe mixture and
constant composition is supported by the data/MC agreement.

\paragraph{Interaction models}
The main conclusion of the present study, the absence of an excess of
muons in data over MC, was obtained for one particular
hadronic-interaction model, QGSJET-II-04. To estimate the effect of the
change of the interaction model on the result, we performed a simplified
study with three other models, namely EPOS-LHC \cite{EPOS-LHC}, SIBYLL~2.1
\cite{SIBYLL-2.1} and QGSJET-01 \cite{QGSJET-01}. The aim of the study was
to estimate the model-related systematic uncertainty of our conclusions.
The simulations were performed with the thinning parameter $\epsilon=
10^{-5}$ and maximal weight restrictions of 40 for hadrons and 4000 for
electromagnetic particles, as suggested by the method of ``optimized
thinning'', Ref.~\cite{Kobal}. For each model, we simulated 400 showers
with $E=10^{17}$~eV and 400 showers with $E=4 \times 10^{17}$~eV for
primary protons and the same amount of showers for primary iron.
The age parameter was determined by averaging of particle densities
in concentrical rings around the shower axis, the muon content of the shower
was studied using the total number of muons, which is uniquely
translate\green{}d \black to muon density by the muon LDF.

It has\begin{table}
\begin{center}
\red
\begin{tabular}{|c|c|c|}
\hline
Model & mean Fe fraction  &  $k$  \\
      &  from $S$, \%     &\\
\hline
QGSJET-II-04 &  57 &0.92$\pm$0.06\\
\hline
EPOS-LHC &     42 & 0.96$\pm$0.06\\
SIBYLL~2.1 &   76 &1.00$\pm$0.07\\
QGSJET-01 &   58 &0.95$\pm$0.06\\
\hline
\end{tabular}
\end{center}
\caption{\label{tab:models}\red
Results for different hadronic-interaction models.
}
\end{table} been shown by an explicit comparison of thinned and
non-thinned showers that the procedure does not change the central values
of LDF-related parameters, though it does introduce additional
fluctuations \cite{livni}. We re-checked that the central values remain
the same as in the full simulation by performing the same simplified
simulations with our principal model, QGSJET-II-04.
Assuming the measured spectrum, we estimated the iron fraction
from $S$, \green repeating the procedure for this case\black. We determined
the coefficient $k$ in the same way as in the main part of the study. The
results are presented in Table~\ref{tab:models}.

We see that though particular best-fit composition differs from model to
model (by $\sim 15\%$ in terms of the iron fraction in the iron-proton
mixture), the systematic uncertainty in $k$, related to the interaction
model, does not exceed $\sim 8\%$. The muon excess is absent in all cases,
in agreement with the observation that the fit never requires primaries
heavier than iron.

\black

\section{Conclusions}
\label{sec:concl}
We have analyzed the \red densities \black of muons ($E_{\mu }>10$~GeV)
registered by underground detectors of the EAS-MSU experiment. Starting
from the Monte-Carlo simulation based on the primary composition inferred
from the surface-detector data alone and on the QGSJET-II-04 hadronic
interaction model, we obtain a good agreement between the simulations and
the data. Assuming that the number of muons in air showers scales with a
coefficient $k$ with respect to the simulation, we constrain \red $k=0.92
\pm 0.0\green 6\black$\black, so that no muon excess ($k>1$) is observed
and $k=1$ agrees with the data at the $90\%$ confidence level. Similar
conclusions are obtained for primary composition assumptions favoured by
the results of other experiments.

We note that the nice agreement between predicted and observed muon
\red densities \black reported here does not necessarily mean that
QGSJET-II-04 gives a correct description of the muon production in any
case. The agreement observed here relates to $E \sim (10^{17}-10^{18})$~eV,
$E_{\mu} \gtrsim 10$~GeV and inner parts of the shower, $r \lesssim
(2-3)R_{0}$. Previous results, collected in Table~\ref{tab:exps},
\begin{table}
\begin{center}
\begin{tabular}{|c|c|c|c|c|c|c|c|}
\hline
Experiment & \red altitude,&\red  $X$, \black& $E$, eV & $E_{\mu}$, &
$r/R_{0}$ &\green$\theta$\black & muon excess\\
& \red m a.s.l.            &\red  g/cm$^{2}$ \black&    & GeV            && &
\red(data over MC)\black \\
\hline HiRes-MIA~\cite{HiRes-MIA} &\red 1500&\red 860 \black&
$10^{17}-10^{18}$ & $\gtrsim
0.85$ & $\gtrsim 10$& N/A & \red yes\\
PAO~\cite{Engel, PAOmu2} &\red 1450&\red 880\black& $\gtrsim
10^{19}$ & $\gtrsim 1$
& $\gtrsim 10$&\green 70$^{\circ}$ & \red yes\\
Yakutsk~\cite{Yak-mu} &\red 100&\red 1020\black& $\gtrsim 10^{19}$ &
$\gtrsim 1$ & $\gtrsim
10$ &\green 45$^{\circ}$ & \red yes\\
IceTop~\cite{IceTop} &\red 2835&\red 680\black& $10^{15}-10^{17}$ &
$\gtrsim {\green 0.2}$ & $\gtrsim
3$&\green 13$^{\circ}$ {\small mean} & \red no\\[3pt]
\hline
\begin{minipage}[c]{2cm}
EAS-MSU\\
(this work)
\end{minipage}  &\red 190&\red 990\black& $10^{17}-10^{18}$ & $\gtrsim 10$
& $\lesssim 3$ &\green 30$^{\circ}$ & \red no\\
\hline
\end{tabular}
\end{center}
\caption{\label{tab:exps}
Comparison with previous studies of the muon excess (see the text for
notations and discussions). }
\end{table}
have been obtained in various different regimes\red, and some at
different altitudes\black. The muon excess reported \red by PAO \black
\green Refs.~\cite{Engel, PAOmu2} \black \red and Yakutsk \black
\cite{Yak-mu} was observed at primary energies $E \gtrsim 10^{19}$~eV and
muon energies $E_{\mu} \gtrsim 1$~GeV. HiRes-MIA~\cite{HiRes-MIA} observed
the excess for $E_{\mu } \gtrsim 0.85$~GeV at $10^{17}$~eV$\lesssim E
\lesssim 10^{18}$~eV. \red Contrary\black, recent preliminary IceTop
results~\cite{IceTop} for GeV muons and $10^{15}$~eV$\lesssim E \lesssim
10^{17}$~eV suggest that no excess is seen. One should not forget also the
important difference between our work and all these studies: here we
investigate the inner parts of EAS, $\red r\black \lesssim (2-3)\, R_{0}$,
while the results of other experiments refer to the outer parts, $\red r
\black \sim 10 \,R_{0}$. Note that at even lower $E$ and higher $E_{\mu}
\gtrsim 1$~TeV, the muon excess may be probed with the help of atmospheric
muons \cite{DedJETPL}, and it has been reported \red in \black
\cite{1504.05853} that QGSJET-II-04 \red seems to \black
\emph{overestimate} the number of muons in this regime. Preliminary
results of the KASCADE-Grande experiment \red (110~m a.s.l., 1022
g/cm$^{2}$) \black at $E\sim 10^{17}$~eV suggest \cite{KASCADE-muon-att}
that the atmospheric attenuation of the muon number \green ($E_{\mu}
\gtrsim 0.23$~GeV, $r/R_{0} \gtrsim 3$) \black is underestimated by all
\red high-energy \black hadronic \red interaction \black models studied
there, including QGSJET-II-04. Clearly, further experimental and
theoretical studies are required to understand the origin of the reported
discrepancies and to arrive at a succesful model of the air-shower
development.

\appendix

\section*{Appendix. Muon number scaling and the surface detectors}
\label{sec:app}
The scaling of the muon number in our Monte-Carlo simulations was
implemented for underground detectors, but
not for the surface-detector observables. This assumption was tested by
the analysis of surface-detector observables in the Monte-Carlo showers
with downscaled number of muons. The downscaling itself is a random
removing of a $(1-k)$ fraction of muons from the CORSIKA showers.
Comparing the showers with $k=k'\equiv0.6$ and $k=1.0$, we found the mean
change in the reconstructed $N_{e}$ of 2.5\%. The root mean
squares of \red the \black variations of the principal observables (primed
quantities correspond to $k=k'$) are: $\sigma((N_{e}'-N_e)/N_e) = 12.5\%$,
$\sigma(s'-s) = 0.047$,
$\sigma(R'-R) = 2.6$~m,
$\sigma(\theta'-\theta) = 0.32^{\circ}$.
We should note that the downscaling of muon number affects the SD
observables stronger than its upscaling, so the presented values of
variations can be considered as upper limits on what we can get with the
upscaling of the muon number. At the same time, these values are smaller
than the experimental uncertainties, see Ref.~\cite{JINST}, therefore we
can neglect the impact of the muon-number scaling on surface-detector
observables.

\section*{Acknowledgements}
ST acknowledges interesting and stimulating discussions of the ``muon
excess'' with Ralf Engel, Sergey Ostapchenko and Gordon Thomson.
Monte-Carlo simulations have been performed at the computer cluster of the
Theoretical Physics Department, Institute for Nuclear Research of
the Russian Academy of Sciences. The experimental work of the EAS-MSU
group \red was funded by \black the Government of the Russian
Federation (agreement 14.B25.31.0010) and the Russian Foundation for Basic
Research (project 14-02-00372). Development of the analysis methods
and application of them to the EAS-MSU data \red were funded by \black the
Russian Science Foundation (grant 14-12-01340).



\end{document}